\documentstyle[12pt,aaspp, flushrt]{article}
\begin{document}
%%%%%%%%%%%%%%%%

\title{X-ray Spectral Survey of WGACAT Quasars,\\
II: Optical and Radio Properties of Quasars \\
	with Low Energy X-ray Cut-offs}

\author{Martin Elvis$^1$, Fabrizio Fiore$^{1,2,3}$, Paolo
Giommi$^3$, \\ Paolo Padovani$^4$}

\affil{$^1$ Harvard-Smithsonian Center for Astrophysics\\
60 Garden St, Cambridge MA 02138, USA}

\affil{$^2$Osservatorio Astronomico di Roma, Monteporzio (Rm), Italy}

\affil{$^3$BeppoSAX Science Data Center, Roma, Italy}

\affil{$^4$Dipartimento di Fisica, II Universit\`a di Roma ``Tor
Vergata'', Via della Ricerca Scientifica 1, I-00133 Roma, Italy}

%\author{\tt (version: 11:45m, 2 June 1997, post-referee) }
%
\begin{abstract}

We have selected quasars with X-ray colors suggestive of a low
energy cut-off, from the ROSAT PSPC pointed archive. We examine
the radio and optical properties of these 13 quasars. Five out of
the seven quasars with good optical spectra show associated
optical absorption lines, with two having high $\Delta$v
candidate systems. Two other cut-off quasars show reddening
associated with the quasar.  We conclude that absorption is
highly likely to be the cause of the X-ray cut-offs, and that the
absorbing material associated with the quasars, not intervening
along the line-of-sight. The suggestion that Gigahertz Peaked
Sources are associated with X-ray cut-offs remains unclear with
this expanded sample.

\end{abstract}


\section{Introduction: Low Energy X-ray Cut-offs in Quasars}
%%%%%%%%%%%%%%%%%%%%%%%%%%%%%%%%%%%%%%%%%%%%%%%%%%%%%%%%%%%%%
\bigskip

The first X-ray spectra of high z quasars showed strong,
unanticipated, low energy cut-offs (Elvis et al., 1994). A
tantalizing connection of these cut-offs with GPS quasars was
also suggested, raising the possibility that these cut-offs were
due to hot, galaxy-scale, medium that also confined the radio
sources. If nuclear absorbing material produces the cut-offs a
different environment must be common at high redshifts. We have
now investigated these cut-offs further.

In the first paper in this series (Fiore et al. 1996, Paper I) we
used the PSPC soft X-ray slope ($\alpha_S$) derived from the
ROSAT pointed PSPC database (using `WGACAT', White, Giommi \&
Angelini 1995) and found the following results:

\noindent
(1) Radio-loud quasars have X-ray colors which differ
($P_{chance}$=1\%) from those of radio-quiet quasars. Most of
these lie in the zone indicating a low energy X-ray cut-off.
In order to affect radio-loud and radio-quiet quasars
differently, these cut-offs must be associated with the quasars,
and not with intervening systems.

\noindent
Spectral fits (Paper I) to the better signal-to-noise PSPC
spectra give more confidence that absorption is at work.  Three
levels of confidence in the presence of a low energy X-ray
cut-off were defined (Table 1): `A': Five quasars require (at
$>$99.9\% confidence) absorption in a power-law plus absorption
model. 3C~109 also shows excess absorption (Allen \& Fabian 1992),
but was excluded from our initial search because it had too large
a Galactic column density. Since it has interesting properties
(\S 2, 4) we include it for comparison with
the other objects. `B': In five more quasars, the fit
suggests absorption but less strongly (at $>$95\% confidence).
`C': In three quasars either absorption or an extremely flat
spectrum ($\alpha <$ 0.25) is required. Absorption is thus a good
working hypothesis to explain the low energy X-ray cut-offs in
this sample of 13 quasars. Absorption only in radio-loud quasars
implies that radio-loud and radio-quiet quasars environments must
differ at high z. Agreement of the results of Paper I with other
ROSAT and ASCA work is good (Paper I).

\noindent
(2) Among the radio-loud quasars, those at high redshift have a
lower mean $\alpha_S$ than those at low redshift (P=0.04\%), due
to those lying in the X-ray `cut-off' zone. A partial correlation
analysis shows that the change in $\alpha_S$ is a redshift, and
not a luminosity, effect. The X-ray cut-offs, and so presumably
the quasar environments, thus show evolution with cosmic epoch.

If the cut-offs are due to photoelectric absorption then the
X-ray column densities of $\sim 10^{22}$~cm$^{-2}$ are implied at
the quasar. The small range of observed values is probably
induced by the spectral range of ROSAT, limiting effective
observed cut-off energies to the range $\sim$0.5-1.5~keV. An
instrument covering both higher and lower energies, with
sufficient sensitivity, is needed to find the true range of
cut-offs in high z quasars.

An absorbing column density of this size will produce
observable effects at other wavelengths. For standard Galactic
dust composition and dust-to-gas ratio (Jenkins \& Savage 1974)
strong reddening ($A_V \sim 5.5$) is predicted. Neutral material
would show strong Hydrogen absorption in a damped Lyman-$\alpha$
system. Ionized material instead will show absorption lines from
other ions, e.g.  CIV, OVI (c.f. NGC~5548, Mathur, Elvis \&
Wilkes 1996), unless the material is highly ionized (or
primordial).

In this paper we seek evidence from optical spectroscopy for or
against absorption in the same sample. We use our positive
findings to investigate the nature of the absorbers.  We also
examine the radio properties of the quasars. We assume a Friedman
cosmology with H$_0$=50~km~s$^{-1}$~Mpc$^{-1}$ and $\Omega$=0.

%%%%%%%%%%%%%%%%%%%%%%%%%%%%%%
\section{\bf Optical Absorption Features}

To search for the predicted optical obscuration we have compiled
data, from the literature, on absorption local to the quasars in
our sample (Table 1), using NED.
\footnote{The NASA/IPAC Extragalactic Database (NED) is
operated by the Jet Propulsion Laboratory, California Institute
of Technology, under contract with the National Aeronautics and
Space Administration.}
Seven of the quasars have optical spectra with adequate
signal-to-noise and resolution to detect absorption lines within
the broad emission lines. Five of these show associated
absorption lines (Table 1), and two more have candidate systems
at high $\Delta$v. Two other quasars have red optical continua
(Table 1). Red continua, and absorption lines close to the
emission line redshift, are quite rare and both suggest absorbing
material near the quasar. Hence, to find one or the other in 9
out of 11 X-ray cut-off quasars, strongly supports the
interpretation of the X-ray cut-offs as due to absorption.

Four of the lowest redshift quasars are from the 3CR radio
catalog. (3C219, once classified as a narrow line radio galaxy, is
now known to have broad H-$\alpha$ and Paschen-$\alpha$ emission
lines, Hill et al., 1996).  Three of the four have red continua
($\alpha_{OUV}\sim 3$, Table 1), which are found in some 15\% of
3CR quasars (Smith \& Spinrad 1980). The binomial probability of
finding 3 red quasars in a sample of 4 by chance is 1\% . We
conclude that X-ray cut-offs and red continua are linked. The red
color of these quasars could be due to an intrinsic spectral shape
(Wills et al., 1995), as in some blazars (Giommi, Ansari \& Micol,
1996). However we have additional evidence that in these objects
the steep slope is almost certainly due to reddening by dust:
3C~219 has Paschen-$\alpha$/H-$\alpha$/H-$\beta$ emission line
ratios that imply $A_V$=1.8$\pm$0.3 (Hill et al., 1996),
consistent with the X-ray column density (Allen \& Fabian 1992);
3C~109 has a polarized optical spectrum that shows broad lines, as
in NGC~1068 , implying $A_V >$ 2.7 (Goodrich \& Cohen, 1984). For
the continuum of 3C~212 to be a reddened version of a normal
quasar requires the similar reddening (A$_V$=2.8) implied from the
observed X-ray cut-off (A$_V\sim$5.5, for a Milky Way dust to gas
ratio, Savage \& Mathis 1979). The X-ray cut-offs in the 3CR
quasars are thus surely due to absorption.

%%%%%%%%%%%%%%%%%%%%
\begin{table}[t]
\caption{\bf Optical Properties of Sample, Redshift Ordered}
%\begin{centering}
\begin{tabular}{|lllllrll|}
\hline
Quasar  &Cut-off &z     &$\alpha_{OUV}$&Assoc.    &$\Delta v$&
Notes& Ref.    \\
        &Class   &      &              &Absorption&(km s$^{-1}$)&
&     \\
\hline
3C219 (0917+45)&C&0.174 &red$^a$  & ---       &  --- &$A_V=1.8$& 1,2 \\
3C109 (0410+11)&A&0.305 &$-$3     &unknown$^a$&      &$A_V>2.7$& 1,3 \\
PKS1334$-$127  &B&0.539 &$-$2.0   &unknown$^a$&---&10\% pol & 4$^b$,5$^b$,6\\
3C207 (0838+13)&A&0.684 &$-$1.83  &OVI,CIV,L$\alpha$&$<$300& & 7,$c$ \\
3C212 (0855+14)&A&1.043 &$-$3.6   & MgII      & $-$1,830  &     & 8,9 \\
S4 0917+624    &C&1.446 &   ---   &unknown$^b$& ---       &     & 10$^b$    \\
S4 0917+449    &B&2.18  &blue$^d$ &unknown$^e$& ---     & --- &  $e$     \\
PKS2351$-$154  &B&2.665 &blue$^d$ &CIV,SiIV,L$\alpha$&$-$1,050& & 11,12 \\
PKS0438$-$436  &A&2.852 &         &Ly$\alpha$?$^b$&$-$8,400&4\% pol.$^f$&13,5
\\
               & &      &         &           &+3,600&  &       \\
               & &      &         &           &+9,300&  &       \\
PKS0537$-$286  &C&3.119 &         &CIV        &$-$4,200  &  & 14    \\
               & &      &         &           &$-$40,000&  &       \\
S4 0636+680    &C&3.174 &         &CIV        &$-$35,000&       & 15    \\
PKS2126$-$158  &A&3.266 &$-$0.5   &CIV, SiIV  &$-$14,900&$A_V<1$& 16,17\\
S4~1745+624    &B&3.889 &$-1.3$   &unknown$^b$&---      & & 18$^b$,19 \\
\hline
\end{tabular}
\smallskip
\footnotesize

$a$. galaxy dominated spectrum.
$b$. insufficient S/N or resolution in optical spectrum.
$c$. absorption lines are present in the Wills et al., (1995) HST spectrum
(S.Mathur 1996, private communication).
$d$. no photometry.
$e$. no published optical spectrum.
$f$. but see Fugmann \& Meisenheimer (1988).

References:
1. Yee \& Oke 1978;
2. Fabbiano et al, 1986, Hill et al., 1996;
3. Rudy et al., 1984, Elvis et al., 1984;
4. Wilkes et al., 1983;
5. Impey \& Tapia 1988;
6. Stickel  \& K\"{u}hr, 1993a;
7. Wills et al., 1995;
8. Smith \& Spinrad 1980;
9. Aldcroft et al., 1994;
10. Stickel \& K\"{u}hr 1993;
11. Roberts et al., 1978;
12. Barthel et al., 1990;
13. Morton et al., 1978;
14. Wright et al., 1978;
15. Sargent et al., 1989;
16. D'Odorico et al., 1997;
17. Kuhn, 1996;
18. Stickel \& K\"{u}hr, 1993b;
19. Hook et al., 1995.

\normalsize

%\end{centering}
\end{table}
%%%%%%%%%%%%%%

For the rest of the sample absorption lines near to the quasar
emission redshift are common (Table 1). When absorption lines lie
within $\Delta v$=5000~km~s$^{-1}$ or so of the emission line
redshift (the criteria are not well-defined) they are likely to
be `associated', i.e. physically connected, with the quasar
(Weyman et al. 1979). Associated absorbers are quite rare, about
20\% of the Weyman et al. sample. Table 1 shows that 4 out of 7
quasars with adequate optical spectra have associated absorbers
with $\Delta v <$5000~km~s$^{-1}$ ($P_{chance}$=0.03\%, using a
binomial probability), and two more at higher $\Delta v$. Hence a
link between X-ray cut-offs and absorbing material is implied for
these high z quasars.  Associated absorbers have been found, via
variability, with $\Delta v$ as high as $-$24,000~km~s$^{-1}$
(Hamman et al., 1997) so the high velocity systems in
PKS~2126$-$158, PKS~0636+680 and PKS~0537$-$286 may also be
physically associated with the quasar.  Given that absorption
exists near the quasars, photoelectric absorption is the
presumptive cause of the X-ray cut-offs.

Since the X-ray absorption is more common at high z (Paper I) the
same should be true of the optical/UV associated
absorbers. Unfortunately associated absorbers are not well
studied below z$\sim$1, where the main UV lines can be studied
only from space.  Studies of Seyfert galaxies hint at a lower
proportion of associated absorbers (Ulrich 1988, G.Kriss 1997,
priv. comm.).

%%%%%%%%%%%%%%%%%%%%%%%%%%%%%%%%%%%%%%%%%%%%%%%%%%%%%%%%%%%%%

\section{Radio Properties}

The radio properties of the quasars are quite varied.  We
collected information from the literature on the radio structure
and spectra of all the sample sources (Table 2, Figure 1) using
the NED database.
The 3C quasars are large ($\sim$100~kpc dia.) classical doubles
with simple, almost pure power-law, spectra (Table 2).  Despite
their large size, the intermediate values of their core-dominance
parameters ($log R_{cl}$ at 5GHz, Table 2) do not indicate an
extreme edge-on orientation ($log R_{cl}< -1.5$, Browne \& Murphy
1987); nor an extreme pole-on orientation, ($log R_{cl}>1.5$).

%%%%%%%%%%%%%%%%%%%%
\begin{table}[t]
\caption{\bf Radio Properties of Sample, Redshift Ordered}
%\begin{centering}
\begin{tabular}{|lllllll|}
\hline
Quasar & z &\multicolumn{2}{c}{Diameter}&$log R_{cl}$& Spectrum$^a$ & Notes \\
       &   &            arcsec& kpc     &        &  &      \\
\hline
3C219 (0917+45)&0.174 &86     &342[1]&$-$1.22[3]&Power Law~$\alpha =-.9$[2]& \\
3C109 (0410+11)&0.305 &47     &280[1]&$-$0.38[4]&Concave~$\alpha
=-(.9-.7)$[2]&\\
PKS1334$-$127  &0.539 &0.0001[5]&0.0008&~~~0.6~[6]&GHz cut-off+MHz excess &
$b$. \\
3C207 (0838+13)&0.684 & 5.5   &  52 [1] &$-$0.15[7]&Concave~$\alpha
=-(.9-.4)$[2]& \\
3C212 (0855+14)&1.043 & 5.1   &  56 [1] &$-$0.31[8]&Power Law~$\alpha =-.8$[2]&
\\
S4 0917+624    &1.446 & 0.021 [8]&0.25  &~~~0.6[9] &GHz cut-off+MHz excess &
   \\
S4 0917+449    &2.18  & 0.019 [10]&0.25  &  &GHz cut-off+MHz excess,& $c$. \\
               &      &       &         &  &+GHz high $\nu$ cut-off &       \\
PKS2351$-$154  &2.665 & $d$   &   ---   &  &           & $c$. \\
PKS0438$-$436  &2.852 & 0.035[13]& 0.47 &1.2[11]&MHz peak?    & $b$. \\
PKS0537$-$286  &3.119 & $d$   &    ---  &1.8[11]&flat, mm cut-off   &       \\
S4~0636+680    &3.174 & 0.0006[13]&0.01 &  &GPS    &       \\
PKS2126$-$158  &3.266 &4.6[12]& 90      &$>$1.1[11]&GPS    & $c$. \\
S4~1745+624    &3.889 & 0.005 [13]&0.07 &  &flat, large MHz excess   [3]   &
   \\
               &      &       &         &  &                   &       \\
\hline
\end{tabular}
\smallskip
\footnotesize
$a$. spectrum references given in caption to figure 1, except as
noted; $b$. mm variable by factors of 2-4 over years (observed
frame, Tornikoski et al., 1996); $c$. core-jet (Marscher A.,
1997, private communication, Neff \& Hutchings 1990); $d$. no
published radio map.

References:
1. Saikia et al., 1995;
2. Herbig \& Readhead 1992;
3. Branson et al., 1972;
4. Riley \& Pooley 1975;
5. Linfield et al., 1990;
6. Browne \& Perley, 1986;
7. Pooley \& Henbest 1974;
8. Jenkins et al., 1977;
9. Xu et al., 1995;
10. Murphy et al., 1993;
11. Browne \& Murphy, 1987;
12. Neff \& Hutchings 1990;
13. Taylor et al., 1994.
\normalsize

%\end{centering}
\end{table}
%%%%%%%%%%%%%%

%%%%FIGURE%%%%
\begin{figure}
\plotone{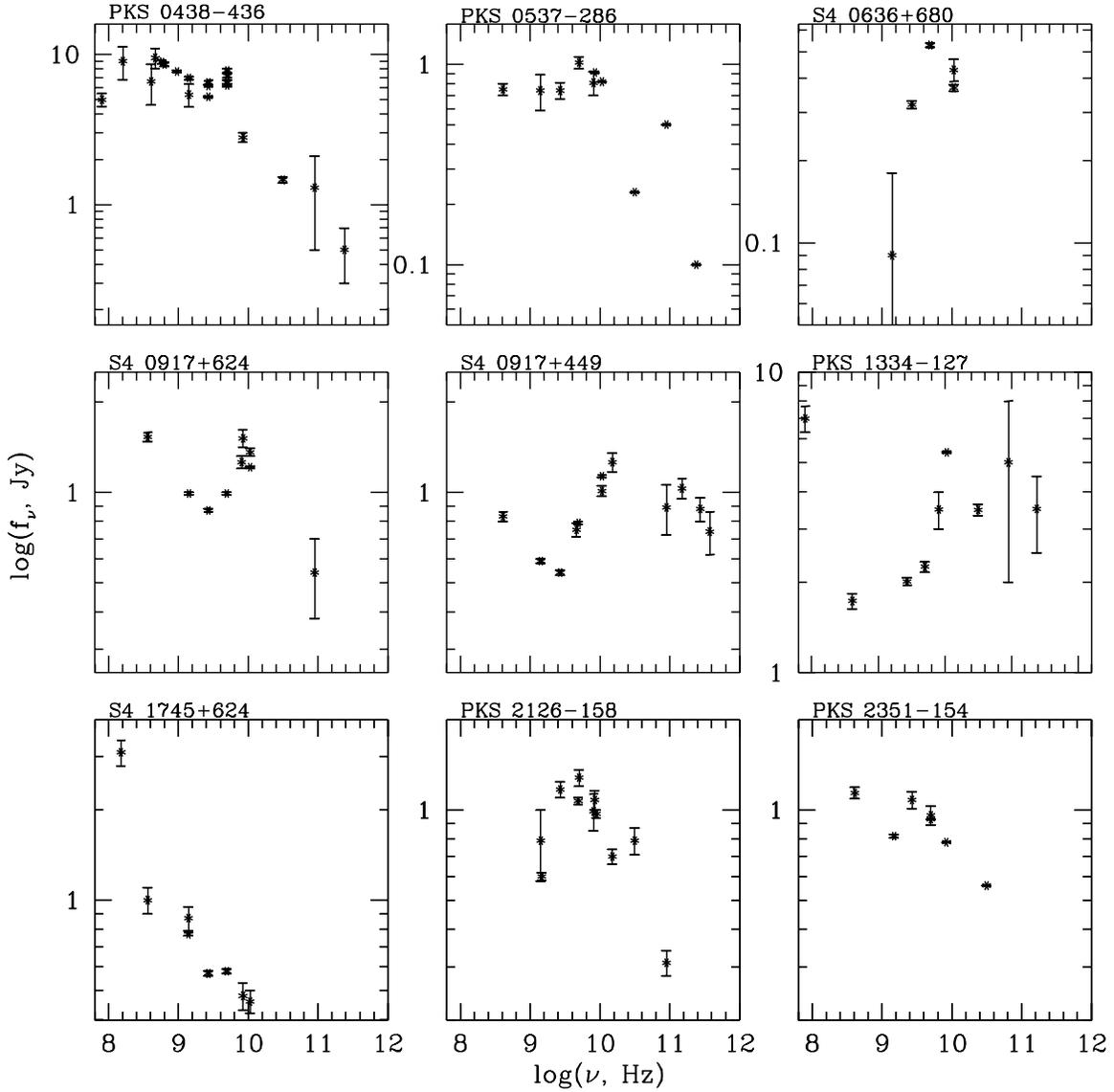}
\caption{Figure 1: Radio spectra of 9 X-ray cut-off sources.
{\em [References -
PKS0438$-$436: Morton et al., 1978, PKSCAT90, K\"{u}hr et al.,
1981, Tornikoski et al., 1996;
PKS0537$-$286: PKSCAT90, K\"{u}hr et al.,1981, Wright et al., 1978,
Tornikoski et al., 1996;
PKS0636+680:  K\"{u}hr et al., 1981;
S4~0917+624:  Douglas et al., 1996, K\"{u}hr et al., 1981, Owen
et al., 1980, Patnaik et al., 1992;
S4~0917+449:  Ficarra et al., 1985, K\"{u}hr et al., 1981, Bloom et
al., 1994;
PKS1334$-$127: PKSCAT90,  K\"{u}hr et al., 1981, Condon et al., 1978,
Baath et al., 1981, Tornikoski et al., 1996;
S4~1745+624: Hales et al., 1990 [6C], White \& Becker 1992, Brundage
et al., 1971, K\"{u}hr et al., 1981, Patnaik et al., 1992, McMahon et
al., 1994;
PKS2351$-$154: PKSCAT90, Quiniento \& Cersosimo 1993, K\"{u}hr et
al., 1981, Neff \& Hutchings 1990;
PKS2126$-$158: Quiniento \& Cersosimo 1993,  K\"{u}hr et al., 1981,
Wright et al., 1991, Steppe et al., 1988, Robson et al., 1985
]}}
\end{figure}
%%%%%%%%%%%%%%

The high z quasars instead are compact and have complex radio
spectra (Figure 1):

\noindent
{\em S4~0636+680 and PKS~2126$-$158} are good Gigahertz peaked source
(GPS) candidates (O'Dea 1990).

\noindent
{\em S4~0917+624, S4~0917+449, PKS~1334$-$127} have evidence for
cut-offs below a few GHz, with lower frequency excesses.
S4~0917+624 appears to be steep at high frequency, and is a
weaker GPS candidate. A lower frequency `excess' suggests a
compact structure surrounded by `relic' larger scale
emission. This in turn would imply repeated outbursts, an
important datum. The latter two sources probably have flat high
frequency spectra, and are likely to be blazars. In particular,
PKS~1334$-$127 shows 3~mm and 1.3~mm band variability, by a
factor of 2 on timescales of a year (observed frame, Tornikoski
et al., 1996), and is optically polarized (Impey \& Tapia 1988).

\noindent
{\em S4~1745+624}, as noted by Stickel (1993), has a steep low
frequency spectrum, with a flat spectrum above 10~GHz (not shown
in figure 1).

\noindent
{\em PKS~0438$-$436, PKS~0537$-$286, and probably PKS~2351$-$154}, have
flat spectra below $\sim$10~GHz and steep ($\alpha\sim -2$)
spectra at higher frequencies, as seen in some blazars (Bloom et
al., 1994). Like PKS~1334$-$127, PKS~0438$-$436 shows 3~mm band
variability, by a factor of 2 on timescales of 3 years (rest
frame, Tornikoski et al., 1996), and is optically polarized
(Impey \& Tapia 1988).

The core-dominance ratio ($log R_{cl}$, Table 2) of the non 3C
quasars are all large ($>$0.5). For normal radio sources this
would strongly indicate quite core-dominated objects, probably
dominated by relativistic beaming. However both Compact Steep
Spectrum (CSS) and Gigahertz Peaked Spectrum (GPS) sources have
large $R_{cl}$ based on 1~arcsec, VLA style, maps. The small size
of their lobes, which dominate their radio emission, leads to a
misleadingly large $R_{cl}$. Closer examination is needed.

The radio survey, NVSS (Condon et al., 1997) provides
polarization measurements. These indicate that 3C~212, S4~0636+680
and PKS~2126$-$158 have low polarization (0.2\%, 0.0\% and 0.3\%,
respectively), and so are likely not to be blazars (March\~{a} et
al. 1996). Four other quasars, S4~0917+449, PKS~2351$-$154 and
S4~1745+624, have high polarization (3.0\%, 1.6\%, 4.6\% and
5.6\%, respectively) and so are probably blazars. At 0.6\%
PKS~0537$-$286 is indeterminate.

Although the limited VLBI data can be ambiguous, three of the
radio sources (S4~0917+449, PKS~2351$-$154, PKS~2126$-$158) may
have a core-jet rather than double lobe geometry, supporting a
blazar interpretation of these sources.
This view is somewhat strengthened by the observation that
PKS~2351$-$154 varied at 1.6$\mu$m (V-band in the rest frame) by
1.1$\pm 0.2$ magnitudes in 4 months (rest frame) and by 0.3$^m$ in
about 1 week (rest frame, Soifer et al., 1983).

%%%%%%%%%%%%%%%%%%%%%%%%%%%%%%%%%%%%%%%%%%%%%%%%%%%%%%%%%%%%%
\section{Discussion}
\bigskip

We have found that a large fraction of the quasars with low
energy X-ray cut-offs and good optical spectra, show absorption
lines and/or reddening, associated with the quasar. We conclude
that photoelectric absorption by the same obscuring material is
the cause of the X-ray cut-offs.

The redshifts of the optical lines relative to the quasar
emission redshift can suggest a site for the
absorbers. Velocities within 2000~km~s$^{-1}$ of the emission
redshift (PKS~2351$-$154, 3C~207, 3C~212, Table 1) can arise due
to motions within a surrounding cluster of galaxies. They could
also be due to material at rest in the host galaxy since the
redshift of the high ionization emission lines typically differ
systematically from the host redshift by about this amount
($<\Delta v>\sim1500~km~s^{-1}$, Tytler \& Fan, 1992).
Larger velocities are most plausibly associated with motions
connected to the active nucleus itself, since the velocities
needed are clearly present in AGN: the widths of the broad
emission lines are 5000-10,000~km~s$^{-1}$, and velocities within
jets approach $v=c$. Absorbers with large $\Delta v$ probably
indicate outflows (PKS~2126-158, PKS~0537$-$286). Good optical
spectra for the rest of the sample would help decide which type
of system is the more common.

The objects divide into at least two groups, which may not share
a common cause for the observed absorption. The high ionization
absorber in the high redshift quasars and in 3C~207 have features
in common. The three other 3C quasars form a separate group of
heavily obscured objects (see \S 2). If the absorbers in the red
3C AGN are unrelated to the rest, then these others will show
even stronger evolution with redshift than derived in Paper~I.

3C~109 resembles a Seyfert 2, in which a nuclear torus, viewed
edge-on, obscures the broad emission line region and continuum
source (e.g. Urry \& Padovani 1995), since 3C~109 has a polarized
optical spectrum showing a normal broad emission line quasar
spectrum (Goodrich \& Cohen, 1992), as does that of the prototype
Seyfert~2 galaxy, NGC~1068 (Antonucci \& Miller 1985).  The 3C
quasars in this sample are all large ($\sim$100~kpc) diameter
radio sources with simple, quasi-power-law spectra, and so could
well be viewed close to edge-on.

The tori in these luminous ($M_V \sim -27$) radio-loud AGN may be
larger than in Seyferts, though. Mathur (1994) notes for 3C~212
that the dust must lie $>$10~pc from the quasar continuum source
in order not to evaporate (T$<$1750~K).  3C~109 and 3C~219 have
similar luminosities and so the dust must be similarly
distant. This distance is about 100 times the broad emission line
region (BELR) radius for this luminosity ($\sim$ 100~light-days,
Kaspi et al. 1996). A closer equivalent situation may be the
ionized absorbers in Seyfert galaxies. These are, marginally,
consistent with being at a similarly large relative distance from
the BELR (Mathur, Elvis \& Wilkes 1996) and so could be due to
the same phenomenon. If however infrared measurements show that
much of the dust is colder, and so more distant (Elvis et al.,
1984), or if absorption line variability is faster in Seyferts
(requiring a denser absorber closer to the continuum source),
then a new type of absorber may be needed for the 3C
absorbers. In 3C~212 the X-ray absorber is partially ionized
(Mathur 1996). Such dusty, ionized absorbers are common in
Seyfert galaxies (Reynolds 1996).

The absorbers in the higher redshift, higher luminosity quasars,
and in 3C~207, all show high ionization. Similar absorbers are
known in Seyfert galaxies (e.g. Mathur, Elvis \& Wilkes 1996).
The high velocity associated absorber in Q~2343+125 (Hamman et
al., 1997) has high ionization, and the authors suggest a link
with the Broad Absorption Line Quasars that may apply here too.
To have significant fractional ionization in CIV ($>10^{-4}$)
requires an ionization parameter, $U$, $<$3 (e.g. Mathur, Elvis
\& Wilkes 1996). With this $U$ the X-ray continuum should recover
to its unabsorbed level at energies below the Oxygen K-edge. The
ionization could be produced either by the quasars' photoionizing
continuum, or from the absorbing medium itself being hot
(T$\sim$10$^7$~K). In these high luminosity quasars ($L_{opt}\sim
50~L(3C273)\sim 10^{13}-10^{14}L_{\odot}$, V\'{e}ron-Cetty \&
V\'{e}ron 1996) a standard quasar spectrum can ionize out to
$r\sim 20~n^{-1/2}L_{13}^{+1/2}U^{-1/2}$~kpc
(where $n$ is the electron density in cm$^{-3}$, and $L_{13}$ is
the luminosity in units of 10$^{13}L_\odot$). This radius is on a
galaxy scale for typical ISM densities, or on a larger scale for
the central densities typical of strong cooling flows ($n$=0.1,
Fabian 1994). If the absorbers have the same densities as the
ionized absorbers in Seyfert galaxies ($n_e\sim
10^{5}-10^{9}$cm$^{-3}$, Mathur, Elvis \& Wilkes 1996) then they
will lie at radii of $\sim$1~pc, a plausible distance. The
ionization state thus allows both nuclear and large-scale
locations for the absorbers.

At least some of the higher redshift quasars show blue continua
(Table 1). PKS2126$-$158 has an accurately power-law-like
spectrum ($\alpha_{OUV}$=$-$0.51$\pm 0.09$, Kuhn 1996). The lack
of curvature in this spectrum requires any reddening (with an SMC
reddening law) to have A$_V<$1.0 (3$\sigma$, Kuhn 1996), a factor
10 lower than is implied by the X-ray cut-off (for a Milky Way
dust-to-gas ratio). A similar ratio is often encountered in
Seyfert galaxies (Reichert et al., 1985, Schachter et al.,
1997). We can invert the argument used above for the red 3C
quasars: If the gas was initially dusty, then heating by the
quasar continuum radiation has probably raised the dust to
temperatures above 1750~K causing dust loss through
sublimation. This implies a maximum distance for the gas from the
continuum source of $\sim$10~pc, and hence $n>$10~cm$^{-2}$. This
supports somewhat a nuclear origin for the absorbers. Any changes
in continuum luminosity will be tracked by nuclear, high density,
absorbers, which offers a diagnostic for their location.

Since the absorbers are common ($\sim$40\% at high z, Paper I)
they must have a large covering factor, unless a special geometry
applies. This is similar to the case of Seyfert galaxies where
X-ray ionized absorbers are seen in at least half the objects
(Reynolds 1997), although only 10\%-25\% show UV absorption lines
(Ulrich 1988, G.Kriss, private communication) probably because
the absorber is too highly ionized.  Using the arguments of
Mathur, Elvis \& Wilkes (1996) this large covering factor would
imply mass loss rates that could be as large as
$\sim$100M$_{\odot}$/year.

The associated absorption lines seen in this sample are mostly of
metal species: OVI, CIV, NV. This implies an enriched absorbing
medium.  The abundances in the absorbers will give a clue to
their origin. Quasar nuclei at $z\sim 3$ often seem to show large
(factor 10) overabundances relative to solar (Hamman \& Ferland
1993). Galaxies at these redshifts are more likely to have
undergone little star formation and so have the low abundances
found in Damped Lyman-$\alpha$ systems (Wolfe et al., 1994). It
will soon be possible to obtain better X-ray and optical spectra
that will allow abundance determinations.

The radio properties of the sample are unclear.  Only two of the
objects have clear Gigahertz Peaked Spectra.  However another
three may show GHz cut-offs with MHz excesses (Table 2, figure
1). Other sources have spectra that are not well defined.  Radio
spectra compiled from the literature can be misleading because of
variability.  O'Dea (1990) found that about half of the radio
loud quasars at high redshift (z$>$3) were GPS candidates based
on a similar literature search of spectra. We conclude that the
tentative connection of X-ray cut-offs with GPS sources (Elvis et
al., 1994) remains ambiguous with this larger sample.
Simultaneous 0.3-30~GHz spectra of all these quasars would
clarify the situation.

If the quasars are dominated by emission from a relativistic jet,
i.e. are blazars, then the picture is quite different: A special
geometry clearly applies, and photoionization by the normal
quasar continuum could well be irrelevant.  The presence of mm
variability in PKS~0438$-$436 and PKS~1334$-$127; of radio
core-jet structures in S4~0917+449, PKS~2351$-$154, and
PKS~2126$-$158; and the large velocities of the absorbers in
PKS~0537$-$286 and PKS~2126$-$158 suggests that the presence of
blazar properties may be related to the absorbers. (Note though
that neither of the mm-variable quasars has known optical
absorption lines.) Ionized absorbers are known in blazar X-ray
spectra (Canizares \& Kruper 1984, Marscher 1988, Madejski et
al., 1991), adn also in the Galactic superluminal sourceds
GRO~J1655$-$40 and GRS~1915+105 (Ueda et al., 1997). If the
absorbers are seen only in end-on jets then the special
orientation required would eliminate the covering factor and mass
loss arguments above.  Good radio spectral and polarization
measurements can decide which of these quasars are blazars. VLA
measurements are scheduled.

If the absorbers are jet related, we might identify the $\Delta
v$ of the ionized absorber as the hot spot advance speed. Similar
hotspot advance speeds are derived in compact steep spectrum
(CSS) quasars (Readhead et al., 1996). (The larger velocities,
$v\sim 0.3c$, required to explain the relativistic beaming
effects in jets may well be much larger than the advance speed of
the hotspot, Pearson 1996.)  The absorber may be the boundary
layer between the jet and the surrounding medium. In this case
the absorber may be shock ionized, rather than photoionized by
the quasar continuum. Entrainment of ambient material in the jet
at the boundary layer (e.g. de Young 1996) may then be probed by
X-ray and optical absorption, allowing study of the entrainment
process and of the composition of the ambient medium.

Although the arguments here favor a nuclear, probably
jet-related, origin for the X-ray absorption found in Paper I, it
is hard in this picture to explain the evolution of this
absorption with redshift . Instead, there is strong evidence that
high z quasars and radio galaxies do lie in a galaxy-scale high
density medium which probably has large column densities. Strong,
highly polarized, Lyman-$\alpha$ emitting gas with knotty complex
morphology is seen aligned with the radio structures - the
`Alignment Effect' (di~Serego Alighieri et al., 1988). Radio
galaxies sometimes show large Lyman-$\alpha$ HI haloes and disks
with a column density of $\sim$10$^{19}$~cm$^{-2}$
(R\"{o}ttgering et al., 1996). In Unified Schemes for AGN
(Barthel 1989, Urry \& Padovani 1995) the central engine would
illuminate this medium in other directions, causing it to be
highly ionized. When we observe a quasar we will be looking
through this medium which will show an X-ray/UV absorber. (The
column density seen will depend on the degree of flattening of
the medium.) In fact CSS quasars show an excess of associated UV
absorbers (Baker \& Hunstead 1996). In PKS~2351$-$154 (and
possibly PKS0537$-$286), in which the absorber $\Delta v$ is
small, the observed X-ray absorption may be from this large scale
medium. Future X-ray observations targetted to this end should
show absorption and so reveal a great deal about this galaxy
scale high z medium.

Better X-ray data are essential. In particular, high resolution
X-ray spectra will be able to measure the redshift of the Oxygen
edge in high redshift quasars. This will test the association
with the optical absorption lines and will directly measure the
ionization state and constrain the abundances in the
absorber. For OVII this edge will appear at $0.7/(1+z)$~keV,
i.e. around 0.2~keV. As an example, the AXAF `Low Energy
Transmission Gratings' (LETGS) have a resolution $E/\Delta
E\sim$1500 at this energy and so can determine an edge velocity
to $\sim 200$~km~s$^{-1}$. The LETGS effective area is
10--20~cm$^{-2}$, requiring long exposure times even on the
brightest high z quasars. Areas of several square meters
(c.f. Elvis \& Fabbiano 1997) are needed to exploit the available
spectral information for a reasonable range of quasar
luminosities at high z.

%%%%%%%%%%%%%%%%%%%%%%%%%%%%%%%%%
\section{Conclusions}

Paper I showed that high redshift radio-loud quasars often show
low energy X-ray cut-offs associated with the quasars, with an
increasing fraction of cut-offs being found at high redshifts. In
this paper we have shown that X-ray cut-offs and optical
absorption associated with the quasar are statistically linked,
and hence that the X-ray cut-offs are indeed due to photoelectric
absorption. A prediction is that lower redshift radio-loud
quasars will have fewer associated UV absorbers.

In contrast to the low z absorbed quasars in the sample, which
are often dusty, the high z absorbers have low dust content and
are highly ionized. The location of the absorber is probably
nuclear, based on the large velocities in some, and on the
absence of dust. The evolution with redshift would be more
naturally explained, however, if the absorbers were on the scale
of the host galaxy or larger.  The absorbers may be associated
with the jets in these sources, given the large number of
core-jet and highly variable radio sources.  In this case the
absorbers may be entrained material, located at the boundary
layer between the jet and the surrounding medium.  Far more work
will be needed to establish this association, however. The
association with GPS radio sources is still ambiguous and
requires better radio data to be confirmed or rejected.

X-ray spectroscopy is the most powerful means of following up
this work: Velocity measurements could clinch the optical
absorber/X-ray absorber association; identifying the relative
ionization states of Oxygen via the O-K edges will reveal the
ionization state of the absorber; combining optical and X-ray
spectroscopy will allow abundance measurements, perhaps
determining the origin of the absorbing material. Other tools
include optical/X-ray variability which could definitively
associate the absorber with the nucleus. Also, searches for a
large scale absorbing medium around high z quasars could improve
the case for a large scale origin in some cases. Radio spectra,
polarization (which are in progress) and VLBI mapping could
definitively determine the nature of the quasars in the sample,
be it GPS, blazar, or another type. Larger samples of cut-off
X-ray quasars can be created using the hitherto unidentified
sources in the ROSAT database.  In intermediate, lower luminosity
quasars, direct X-ray imaging could search for extended X-ray hot
atmospheres.  X-rays can thus give us a new set of tools to
investigate the high z quasar environment.

%%%%%%%
\acknowledgments
%\vspace{-0.1in}

We thank Smita Mathur, Niel Brandt, Alan Marscher, Chris O'Dea
Gerry Kriss and Jonathan McDowell for valuable discussions.  This
work was supported in part by NASA contract NAS8-39073 (ASC), and
NASA grants NAGW-2201 (LTSA) and NAGW5-3066 (ADP).  This research
has made use of data obtained through the High Energy
Astrophysics Science Archive Research Center Online Service ,
provided by the NASA-Goddard Space Flight Center.  This research
has made use of the NASA/IPAC Extragalactic Database (NED) which
is operated by the Jet Propulsion Laboratory, Caltech, under
contract with the National Aeronautics and Space
Administration. This research has made use of the BROWSE program
developed by the ESA/EXOSAT Observatory and by NASA/HEASARC.

%%%%%%%

%%%%%%%%%%%%%%
\end{document}